\documentclass[prl,twocolumn,preprintnumbers,groupedaddress,amsmath,amssymb, longbibliography]{revtex4-1}
\usepackage{graphicx}
\usepackage{mathrsfs}
\usepackage{amsfonts}
\usepackage{times}
\usepackage{amsmath}
\usepackage{leftidx}
\usepackage{tikz}
\usepackage{bbold}
\usepackage{braket}
\usepackage{mathtools}
\usepackage[colorlinks,linkcolor=blue,citecolor=blue]{hyperref}

\newcommand{\black}[1]{\textcolor{black}{#1}}

\newcommand{\Psith}{\Psi_{\theta}}
\newcommand{\ham}{\mathcal{H}}
\newcommand{\dtau}{\delta_{\tau}}

\begin{document}
\title{Purifying Deep Boltzmann Machines for Thermal Quantum States}

\author{Yusuke Nomura$^1$}
\email[These two authors contribute equally to the paper. E-mail at: ]{yusuke.nomura@riken.jp}
\affiliation{$^1$RIKEN Center for Emergent Matter Science, 2-1 Hirosawa, Wako, Saitama 351-0198, Japan}

\author{Nobuyuki Yoshioka$^{2,3}$}
\email[These two authors contribute equally to the paper. E-mail at: ]{nyoshioka@ap.t.u-tokyo.ac.jp}
\author{Franco Nori$^{3,4,5}$}
\affiliation{$^2$Department of Applied Physics, University of Tokyo, 7-3-1 Hongo, Bunkyo-ku, Tokyo 113-8656, Japan}
\affiliation{$^3$Theoretical Quantum Physics Laboratory, RIKEN Cluster for Pioneering Research (CPR), Wako-shi, Saitama 351-0198, Japan}
\affiliation{$^4$RIKEN Center for Quantum Computing (RQC), Wako-shi, Saitama 351-0198, Japan}
\affiliation{$^5$Physics Department, University of Michigan, Ann Arbor, Michigan 48109-1040, USA}

\date{\today}

\begin{abstract}
We develop two cutting-edge approaches to construct deep neural networks representing the purified finite-temperature states of quantum many-body systems.
Both methods commonly aim to represent the Gibbs state by a highly expressive neural-network wave function, exemplifying the idea of purification.
The first method is an entirely deterministic approach to generate deep Boltzmann machines representing the purified Gibbs state exactly. 
This strongly assures the remarkable flexibility of the ansatz which can fully exploit the quantum-to-classical mapping.
The second method employs stochastic sampling to optimize the network parameters such that the imaginary time evolution is well approximated within the expressibility of neural networks.
Numerical demonstrations for transverse-field Ising models and Heisenberg models show that our methods are powerful enough to investigate the finite-temperature properties of strongly correlated quantum many-body systems, even when the problematic effect of frustration is present.
\end{abstract}
\maketitle

{\it Introduction.---} 
The thermal behavior of quantum many-body systems is one of the most fundamental problems in physics. 
Statistical mechanics states that the density matrix describing a system in thermal equilibrium, governed by a Hamiltonian $\mathcal{H}$ at an inverse temperature $\beta$, is given by the Gibbs state  $\rho = e^{-\beta \mathcal{H}}/{\rm Tr}[e^{-\beta \mathcal{H}}]$. 
Computing and extracting physical properties from the Gibbs state is a significant challenge 
to understand natural phenomena, which in reality all occur at finite $\beta$.

One of the most celebrated numerical techniques in lattice systems is the quantum Monte Carlo (QMC) method~\cite{suzuki_1976, suzuki_1977, sandvik_1999, gubernatis_2016}, typically based on the path integral formalism of the partition function.
The QMC method yields numerically exact solutions when the positive definiteness is assured; otherwise, the infamous negative sign problem arises.
Many physically intriguing system falls into the latter category, and therefore various efforts have been devoted to overcome this difficulty: tensor-network-based algorithms ~\cite{verstraete_ft_2004, zwolak_2004, feiguin_2005, white_2009} mostly applied to one-dimensional (1D) systems, dynamical mean-field theory~\cite{georges_1996} which becomes exact in the infinite coordination-number limit, diagrammatic Monte Carlo methods~\cite{prokofev_1998, houcke_2010}, to name a few~\cite{motta_2019, irikura_2020, iitaka_2020}.
In another notable approach~\cite{sugiura_2012, sugiura_2013, takai_2016, iwaki_2020} using the thermal pure quantum (TPQ) states, one can extract the ensemble property from a single pure state that represents the thermal equilibrium.
We point out, however, that it remains extremely challenging to establish a methodology that is both reliable and scalable for finite-temperature calculations in two-dimensional (2D) systems---the most exotic and intriguing realm in quantum many-body problems.

Neural networks, initially developed for classical data processing in the context of machine learning, offer a very strong methodology for quantum physics~\cite{carleo_2017, torlai_tomography_2018, wang_2018, carleo_2019, melko_2019, melkani_2020, ahmed_2020}. 
As was firstly demonstrated by Carleo and Troyer~\cite{carleo_2017}, neural networks applied as variational wave functions, commonly dubbed as the neural quantum states, are capable of simulating ground states~\cite{carleo_2017, nomura_2017, saito_2017, cai_2018}, excited states~\cite{choo_2018, nomura_2020, yoshioka_solid_2021}, and even out-of-equilibrium property~\cite{carleo_2017, schmitt_2020, xie_2021, yoshioka_2019, nagy_2019, hartmann_2019, vicentini_2019, yuan_2020} of strongly correlated systems up to unprecedently large size.
Among the tremendous variety of network structures, Boltzmann machines with restricted connectivity, known to be universal approximators for arbitrary real/complex amplitudes~\cite{roux_2008,guido_2011}, are useful for statistical mechanics and quantum information.
The Boltzmann machines with the shallowest structure are already powerful enough to compactly express complex quantum states with extensively-growing quantum entanglement~\cite{deng_prx_2017, levine_2019}. 
Furthermore, deep Boltzmann machines (DBMs), i.e., the ones with multiple hidden layers, are guaranteed to provide an efficient description for an even wider range of quantum states~\cite{gao_2017}.
Strongly motivated by their extremely high representability, the ground states in quantum many-body spin systems have been successfully simulated by DBMs~\cite{carleo_2018}.

In the present study, we provide two state-of-the-art methods to construct DBMs that capture the finite-temperature behavior of quantum many-body systems.  
Both methods share the strategy of employing DBMs to express the purified Gibbs state. 
Namely, a mixed state under imaginary time evolution is compactly encoded as a pure DBM wave function in the enlarged Hilbert space. 
In the first method, we find a completely deterministic way to construct DBMs, realizing the exact purified expression of finite temperature states. 
This proves the remarkable flexibility and power of DBMs for investigating finite-temperature many-body phenomena.
In the second method, we provide a stochastic way to simulate the imaginary time evolution which exploits the versatile expressive power of DBMs as approximators.
For demonstration, we apply these methods to the 1D transverse-field Ising (TFI) model
and the 2D $J_1$--$J_2$ Heisenberg model on the square lattice to find surprisingly high accuracy compared to numerically exact methods.
We emphasize that only a polynomial number of auxiliary spins suffices in both approaches, yielding a huge computational advantage even under the presence of the problematic effect of frustration.

\begin{figure}[tb]
%\vspace{0cm}
\begin{center}
\includegraphics[width=0.5\textwidth]{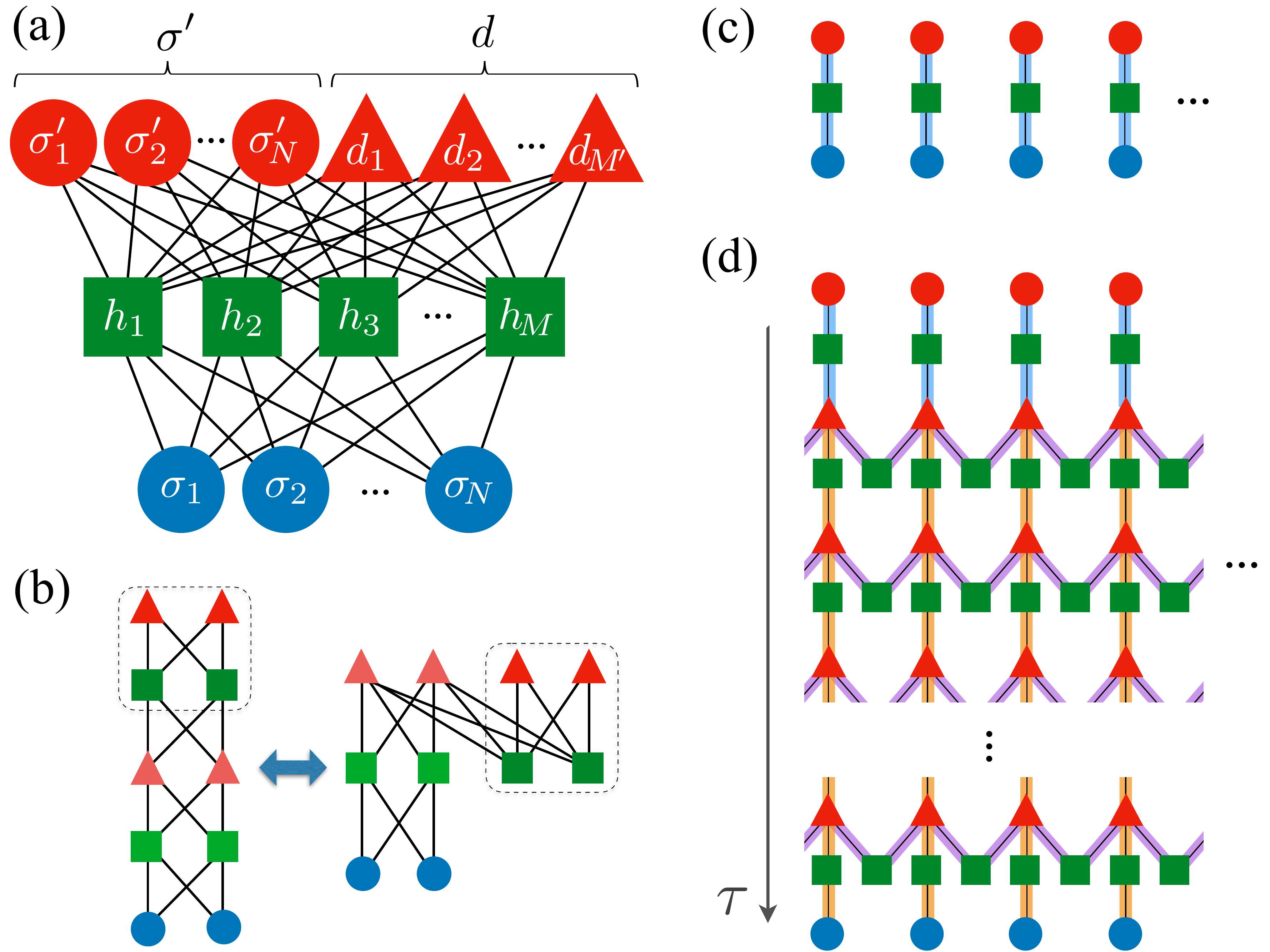}
%\vspace{0.0cm}
\caption{
(a) Structure of the three-layer DBM used in the present study. 
The visible layer (blue) corresponds to physical spins $\sigma_i$. 
Two hidden layers are distinguished as hidden (green) and deep (red) layers;
while the hidden layer simply consists of hidden spins $h_j$, the deep layer is composed of the deep spins $d_k$ and ancilla spins $\sigma_i'$, which are introduced to purify the Gibbs-state density matrix. 
Here, the number of physical/ancilla, hidden, and deep spins are denoted as $N$, $M$, $M'$, respectively.
(b) Two different ways of depicting the identical DBM structure.
A network with seemingly many hidden layers can always be recast into that with only two hidden layers. 
(c) The DBM representing the infinite-temperature state $\ket{\Psi(T=\infty)}$  for quantum spin-1/2 systems.  
(d) The DBM construction for the finite-temperature states of the 1D TFI model 
[$ | \Psi( T ) \rangle $ in Eq.~(\ref{eq:psi_finiteT})].
The arrow denotes the growth of the DBM structure along the imaginary-time $\tau$ propagation.
Light blue, orange, and purple bonds in (c) and (d) have couplings $\mathrm{i} \frac{\pi}{4}$, $\frac{1}{2} {\rm arcosh} \left( \frac{1}{{ \rm tanh} ( \Gamma \delta_\tau )}\right ) $, and $\frac{1}{2}{\rm arcosh} (e^{2 J \delta_\tau }) $, respectively. }
\label{fig:DBM}
\end{center}
\end{figure}

{\it DBM for purification.---} 
Throughout this work, we utilize the idea of purification to represent the Gibbs state. 
Namely, the finite-temperature density matrix $\rho$ of a target system $\mathcal{S}$ is encoded as a pure state in an extended system $\mathcal{S} + \mathcal{A}$, such that $\rho$ is obtained by tracing out the ancillary system $\mathcal{A}$~\cite{nielsen_2000}. 
For instance, the purification of an infinite-temperature state can be generated by the superposition  $\sum_{x} \ket{x}_{\mathcal{S}} \ket{a_x}_{\mathcal{A}}$, where $\{\ket{x}_{\mathcal{S}}\}$ is the complete orthonormal basis set of the target system, while $\{\ket{a_x}_{\mathcal{A}}\}$ is an orthonormal but not necessarily complete basis set of ancillary system.

For the sake of concreteness, let us consider a quantum many-body spin-1/2 system. 
We introduce $N_{\rm site}$ binary degrees of freedom $\{\sigma_i\}$ so that $\ket{\sigma} = \ket{\sigma_1, ..., \sigma_{N_{\rm site}}}$ spans the Hilbert space of the target system $\mathcal{S}$. Hereafter we call them physical spins.
As the ancillary system $\mathcal{A}$, we introduce an identical number of ancilla spins $\{\sigma_i'\}$ so that an arbitrary mixed state can be purified in principle.
While the purification of a mixed state is not unique, here we exclusively take the purified infinite-temperature state as
$ | \Psi(T \! =\! \infty) \rangle =  \bigotimes_{i=1}^{N_{\rm site}} \left( | \uparrow \downarrow^\prime \rangle  +  | \downarrow \uparrow^\prime  \rangle \right)_i $, and perform the imaginary time evolution as $\ket{\Psi (T)} = e^{-\beta \mathcal{H}/2}\otimes \mathbb{1}' \ket{\Psi(T=\infty)}$, with $\beta=1/T$ to simulate the finite-temperature Gibbs state.
Note that the infinite-temperature state $\rho_{\infty} = \mathbb{1}/2^{N_{\rm site}}$ and the finite-temperature state $\rho_T = e^{-\beta \mathcal{H}}/{\rm Tr}[e^{-\beta \mathcal{H}}]$ are reproduced by tracing out the ancilla spins.

Intriguingly, the purified Gibbs state at each temperature can be efficiently expressed by the DBM. 
In particular, we use the DBM with two hidden layers [see  Fig.~\ref{fig:DBM}(a)] to represent a purified wave function whose amplitude $\Psi(\sigma, \sigma') = \braket{\sigma, \sigma'|\Psi}$ is parametrized as 
\begin{eqnarray}
\Psi (\sigma, \sigma') &=& \sum_{h, d} \phi(\sigma, \sigma'; h, d), \label{eqn:DBMansatz}\\
\phi(\sigma, \sigma'; h, d) &=& 
\exp  \biggl[ \sum_{j} b_{j}h_{j} \! +  
\! \sum_{ji} \!  h_{j}  ( W_{ji} \sigma_{i} \! + 
\! W_{ji}^{\prime} \sigma^{\prime}_{i} ) \!  \nonumber\\
&&  + \!  \sum_{jk} W_{jk}^{\prime} h_j d_{k} \biggr],
\label{eq:psidbm}
\end{eqnarray}
where we have introduced hidden spins $\{h_j\}$ 
and deep spins $\{d_k\}$,
in addition to the physical spins $\{\sigma_i\}$ constituting the visible layer and ancilla spins $\{\sigma_i'\}$ allocated in the second hidden layer (we define all spins to be either $+1$ or $-1$).
This structure is ``universal" in terms of the DBM architecture; the arbitrary multi-hidden layer structure can be rearranged to have only two hidden layers as shown in Fig~\ref{fig:DBM}(b).
The number of complex variational parameters 
$\mathcal{W} = \{ b,W,W^{\prime} \}$ \footnote{The renormalization factor is omitted in the expression. 
We ignore bias terms for visible and deep units because they are irrelevant in the present case.}
are directly related to the number of $h$ and $d$ spins, which controls the representability of the DBM wave function. 
Note that the purification technique for neural networks has been considered in the context of quantum tomography~\cite{torlai_NDO_2018} and dissipative quantum physics~\cite{yoshioka_2019, nagy_2019, hartmann_2019, vicentini_2019}.

{\it Method (I): Analytic purification using DBM.---} 
%Now we concretely show how to construct DBMs that exactly express Gibbs states. 
In Method (I), we analytically construct DBMs that exactly reproduce the behaviours of Gibbs states.
First, to represent the infinite-temperature state, we introduce a DBM with $N_{\rm site}$ hidden and ancilla spins [Fig.~\ref{fig:DBM}(c)].
By setting the parameters $\mathcal{W}_{\infty}$ as $b_j =0$ and $W_{ji} = W'_{ji} = \mathrm{i} \frac{\pi}{4} \delta_{ji}$, we find that the analytical expression of the DBM wave function becomes $\Psi(\sigma,\sigma')   =   \prod_i 2 \cosh  \left[ \mathrm{i} \frac{\pi}{4} ( \sigma_i +  \sigma^{\prime}_{i} )  \right ]$.
This exactly reproduces the $ | \Psi(T \! =\! \infty) \rangle$ described above. 

Next, to express finite-temperature states $ | \Psi(T) \rangle$ analytically, we introduce the Suzuki-Trotter decomposition~\cite{suzuki_1976}. 
\textcolor{black}{Namely, given the Hamiltonian $\mathcal{H} = \sum_{\nu} \mathcal{H}_{\nu}$, the purified Gibbs state up to the Trotter error is expressed as}
\begin{eqnarray}
 | \Psi( T ) \rangle :=  \left( \left[ \prod_{\nu} e^{-\delta_{\tau}\mathcal{H}_{\nu}}  \right] ^{ N_{\tau} } \otimes \mathbb{1}' \right)
 \ket{\Psi( T \!  = \! \infty)},
\label{eq:psi_finiteT}
\end{eqnarray}
\textcolor{black}{where $\delta_\tau = \beta/ 2 N_{\tau}$ is the propagation time step. }
%expressed as the operation of each propagation operator $e^{-\dtau \ham_k}$ for  can be represented by the DBM.
\textcolor{black}{Remarkably, thanks to the flexible representability of the DBM, we can find analytical solutions for Eq.~\eqref{eq:psi_finiteT}: Starting from the initial DBM state $| \Psi( T \!  = \! \infty) \rangle$ given by the $\mathcal{W}_\infty$ parameters, each short-time propagation $e^{-\dtau \ham_{\nu}}$ can be implemented exactly by modifying the parameters and structure of the DBM}~\footnote{For spin-1/2 systems, imaginary-time propagators can be analytically represented by the DBM. This has been discussed in the context of classical many-body interacting systems as well~\cite{yoshioka_transforming_2019}.},
\textcolor{black}{whose the explicit expression depends on the form of the Hamiltonian}.
%The corresponding DBM state gives the finite-temperature Gibbs state up to the Trotter error. 

As a concrete example, let us consider the TFI model on the $N_{\rm site}$-spin chain under periodic boundary condition. 
The Hamiltonian is given by ${\mathcal H} = {\mathcal H_1} +{\mathcal H_2}$, with ${\mathcal H_1} = - J \sum_i \sigma^z_i \sigma^z_{i+1} $ and ${\mathcal H_2} = - \Gamma \sum_i \sigma^x_i $, where $\sigma_i^a (a=x, y, z)$ denotes the Pauli matrix operating on the $i$-th site. 
We take the Ising-type interaction $J$ as the energy unit ($J=1$), and $\Gamma$ as the strength of the transverse magnetic field. 
Using Method (I), we can analytically construct the finite-temperature state using the DBM.
Solutions to realize the propagation
by updating the DBM parameters from ${\mathcal W}$ to $\bar{\mathcal W}$, i.e.,  $e^{-\delta_{\tau} \mathcal{H}_{\nu}}  | \Psi_{\mathcal W} \rangle = C | \Psi_{\bar{\mathcal W}} \rangle $ ($C$ : a constant) can be sketched as follows~\cite{carleo_2018}:
\textcolor{black}{For $\nu=1$ (interaction propagator), we add a single hidden spin for each neighboring visible spins $\sigma_i$ and $\sigma_{i+1}$.
For $\nu = 2$ (transverse-field propagator), we add new hidden- and deep-spin layers between the visible and  neighboring hidden layers. 
All the DBM parameters are determined analytically [see Supplementary Materials (SM) for detailed implementation~\footnote{See Supplementary Materials for more details of the method (URL to be added).}].}
As a consequence of propagarions, the DBM architecture grows as in Fig.~\ref{fig:DBM}(d),
and the number of hidden and deep spins scale as $\mathcal{O}(N_{\tau} N_{\rm site})$,
\textcolor{black}{
which is also common among general local Hamiltonians.
}

\textcolor{black}{Now that the Gibbs states are represented, let us discuss how the physical quantities are computed using the DBM framework in general. }
Since the expectation value of a physical observable $\langle {\mathcal O} \rangle = \frac{   \langle \Psi( T ) |{\mathcal O} \otimes {\mathbb 1 }' | \Psi( T ) \rangle }{ \langle \Psi( T ) |  \Psi( T ) \rangle }$ becomes analytically intractable, we use the Monte Carlo (MC) method for its numerical estimation.
The sampling weight is based on the expression of the normalization factor of the DBM state [Eq.~\eqref{eqn:DBMansatz}] given as 
\begin{eqnarray}
&&\langle \Psi |  \Psi \rangle = \sum_{\sigma, \sigma'} \bigl | \Psi(\sigma,\sigma') \bigr |^2 \nonumber \\
&&\ \ \ =  \sum_{\sigma, \sigma'} \sum_{h_1,d_1,h_2,d_2} \phi^{\ast}(\sigma,\sigma'; h_1, d_1) \phi(\sigma,\sigma'; h_2, d_2).
\end{eqnarray}
Namely, we sample over the configurations of $(\sigma, \sigma', h_1, h_2, d_1, d_2)$ weighted by the product of amplitudes as $w (\sigma, \sigma'; h,  d )  \equiv   \phi^*(\sigma,  \sigma'; h_1,  d_1)~\phi(\sigma, \sigma'; h_2,  d_2) $.
Alternatively, it is possible to trace out the hidden spins $h$  analytically, and use $\sum_{h_1, h_2} \phi^* (\sigma, \sigma'; h_1, d_1) \phi(\sigma, \sigma';h_2, d_2)$ as the MC sampling weight over configurations of $(\sigma, \sigma', d_1, d_2)$ (or trace out $d$ spins and sample over $h$ spins).
See SM for more details of the method~\cite{Note3}.

\begin{figure}[tb]
%\vspace{0cm}
\begin{center}
\includegraphics[width=0.48\textwidth]{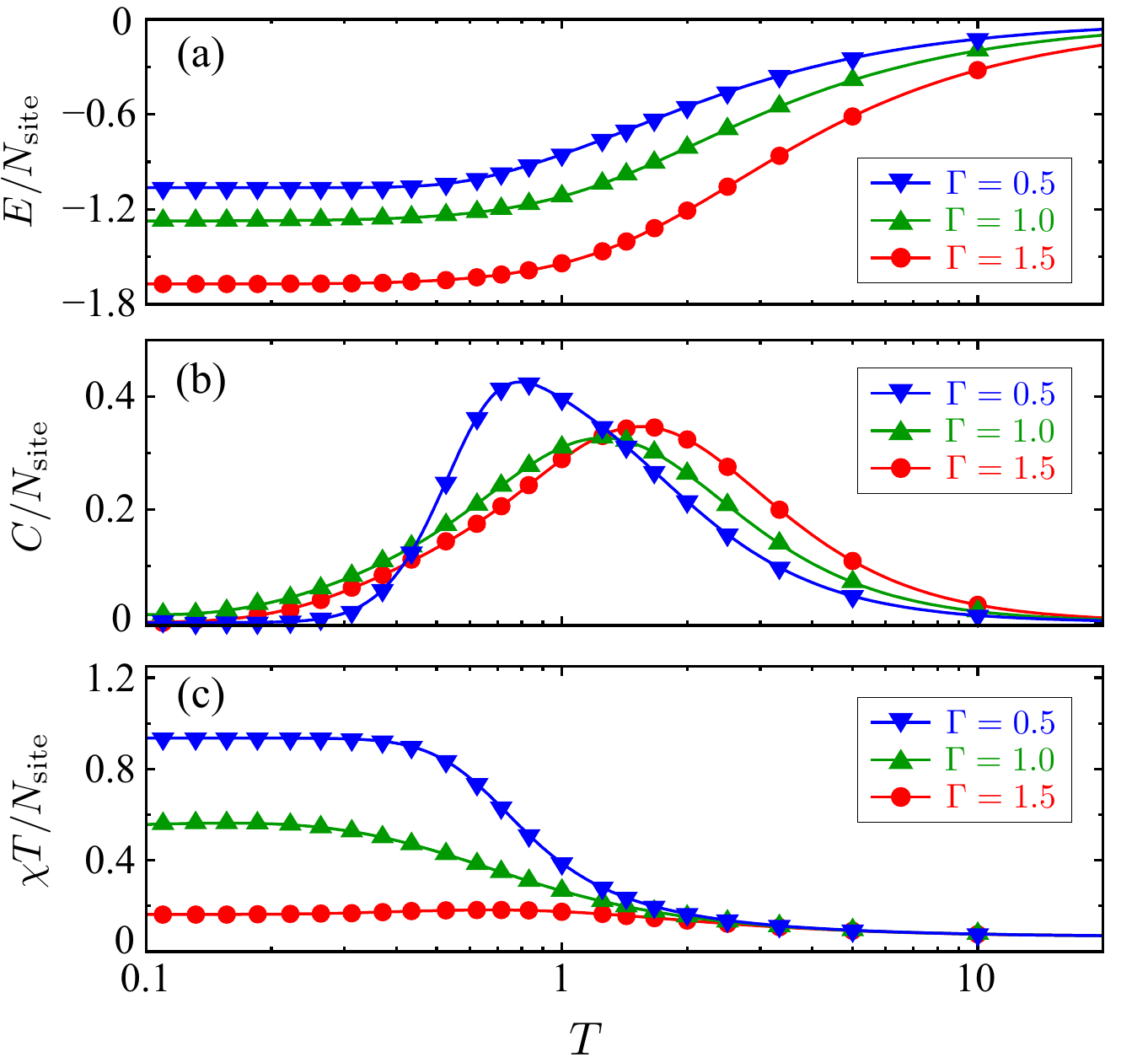}
%\vspace{0.0cm}
\caption{
Finite-temperature calculations for the 1D TFI model on a 16-site chain ($N_{\rm site}=16$) with periodic boundary condition: (a) Energy $E$, (b) specific heat $C$, and (c) susceptibility $\chi = \frac{1}{T} \sum_i \langle \sigma_0 \sigma_i \rangle $. The symbols denote the DBM results [Method (I) with $\delta_\tau = 0.05$], which agree well with the exact-diagonalization results (solid curves). }
\label{fig:TFI}
\end{center}
\end{figure}

\textcolor{black}{To verify the construction of the DBM and the proposed MC sampling strategy,} we apply the method to the 16-site TFI model. 
Figure~\ref{fig:TFI} shows the DBM results (symbols) for the temperature dependence of the (a) energy, (b) specific heat, and (c) susceptibility. 
As expected, the DBM results follow the exact temperature evolution (solid curves). 
This confirms the remarkable representability of the DBM not only at zero temperature~\cite{carleo_2018} but also at finite temperatures, offering an intriguing quantum-to-classical mapping.

{\it Method (II): Numerical purification using DBM.---} 
When the weight of individual spin configuration can be taken to be always positive, Method (I) is quite useful and provides numerically exact finite-temperature results. 
However, when the frustration exists in the spin Hamiltonian, for instance, we cannot avoid the existence of negative weights as in other finite-temperature calculations based on the QMC method. 
While it is possible to construct finite-temperature states analytically, the estimation of physical quantities becomes extremely difficult because of the negative sign problem. 

To make the application to frustrated models possible, we propose an alternative method which employs DBMs with only ancilla spins $\sigma'$ in the second hidden layer.
By tracing out hidden spins $h$, such a purified DBM wave function has a simple form: 
\begin{eqnarray}
\Psi(\sigma,\sigma') =  \prod_j 2 \cosh  \Bigl[ b_j \! + \!  \sum_i ( W_{ji}  \sigma_i  \! +\!  W'_{ji}  \sigma_i^{\prime} )  \Bigr ].
\label{eq:simleDBM}
\end{eqnarray}
Then, we can avoid negative signs by simply employing $\bigl |\Psi(\sigma,\sigma') \bigr|^2$ as the weight for sampling over $\sigma$ and $\sigma'$ spins. 
However, in this case, differently from Method (I), the imaginary-time evolution cannot be followed analytically. 
Instead, we need to update parameters numerically at each time so that the DBM obeys the imaginary-time evolution starting from the infinite-temperature state $\Psi_{\infty}(\sigma,\sigma') = \prod_i 2 \cosh  \left[  \mathrm{i} \frac{\pi}{4} ( \sigma_i +  \sigma^{\prime}_{i} )  \right ]$ (recall that this infinite-temperature DBM does not require $d$ spins). 
For this purpose, we employ the stochastic reconfiguration (SR) method~\cite{sorella_1998, sorella_2001, Note3} \footnote{The same optimization scheme has been developed independently in artificial-intelligence community, called ``natural gradient''~\cite{amari_1992,amari_1998}.}.
\textcolor{black}{The SR optimization is designed to minimize the distance between quantum states following the exact and variational imaginary-time evolution, as much as possible within the expressive power of the DBM wave function in Eq.~(\ref{eq:simleDBM}).}
The expressive power is systematically controlled by the number of $h$ spins; 
It is ensured that any quantum states can be represented exactly (universal approximation) by an infinitely wide network structure~\cite{roux_2008,guido_2011}. 
%\textcolor{black}{Note that the accuracy can be further improved by imposing symmetry on the wave function.}

\begin{figure}[tb]
%\vspace{0cm}
\begin{center}
\includegraphics[width=0.48\textwidth]{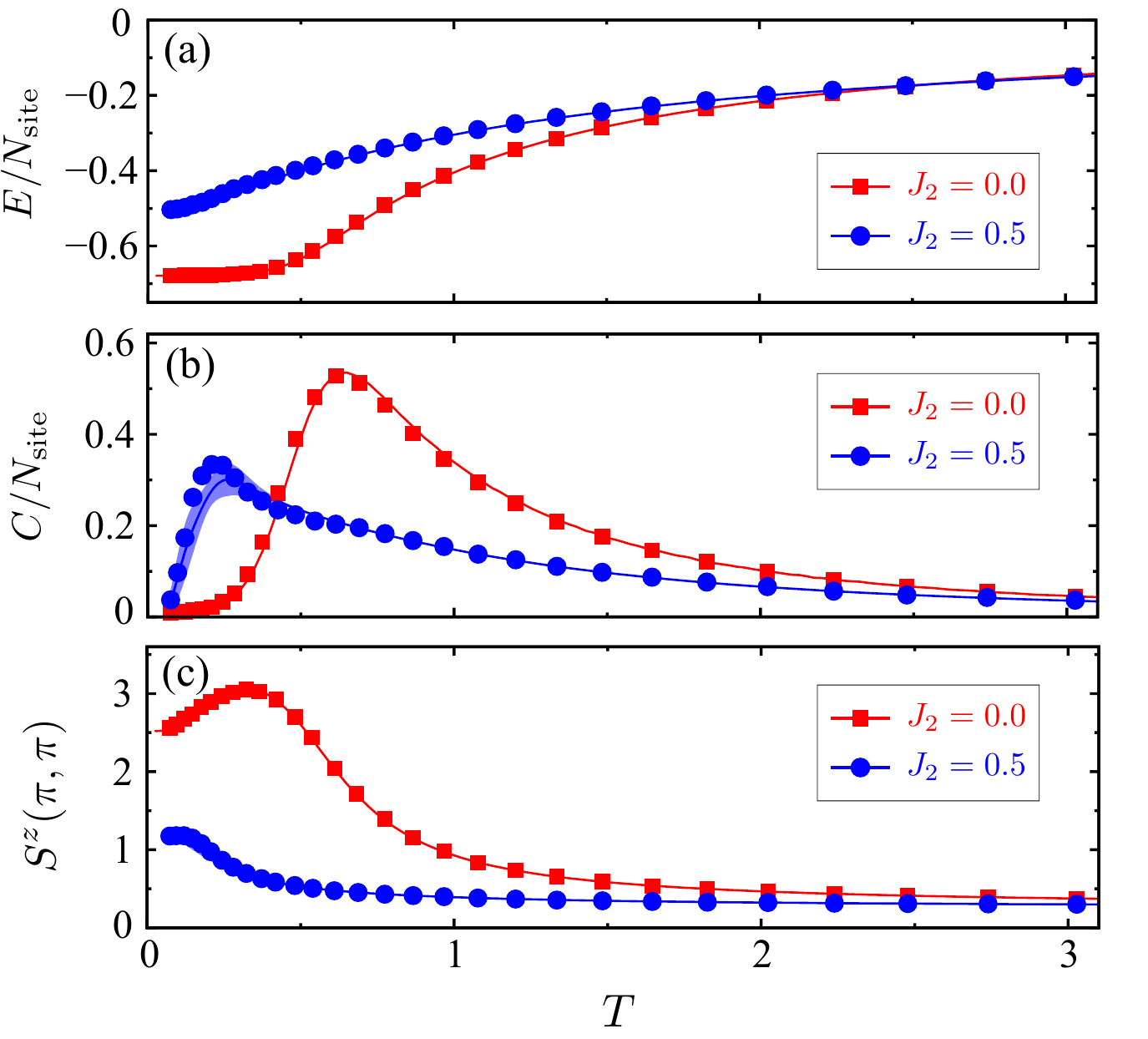}
%\vspace{0cm}
\caption{ 
Finite-temperature calculations for the 2D $J_1$-$J_2$ Heisenberg model ($J_1=1$) on the $6\times6$ square lattice with periodic boundary condition: (a) Energy $E$, (b) specific heat $C$, and (c) $z$ component of the spin structure factor $S^z({\bf q}) = \frac{1}{N_{\rm site}} \sum_{ij} e^{i {\bf q} \cdot( {\bf R}_i - {\bf R}_j) } \langle S^z_i S^z_j \rangle $ at ${\bf q} = (\pi,\pi)$. 
The symbols denote Method (II) results with $8N_{\rm site}$ hidden spins, which show a good agreement with the numerically exact references (solid curves) obtained by Method (I) with $\delta_\tau = 0.005 $ ($J_2=0$) and the TPQ method ($J_2=0.5$). 
The TPQ calculations are performed with ${\mathcal H}\Phi$~\cite{kawamura_2017}. 
The shaded regions show the size of the error bars of the TPQ results [in Method (I), the size of the error bars is small].}
\label{fig:J1J2}
\end{center}
\end{figure}

%{\it Results for Method (II).---} 
\textcolor{black}{As a demonstration of Method (II)}, we turn to a highly challenging problem: frustrated spin systems. 
As a representative, here, we consider the 2D antiferromagnetic $J_1$--$J_2$ Heisenberg model on the $L\times L$ square lattices with periodic boundary condition.
The Hamiltonian reads
${\mathcal H} =  J_1 \sum_{ \langle i, j \rangle}  {\bf S}_i  \cdot {\bf S}_j +  J_2 \sum_{ \langle \langle  i, j \rangle \rangle}  {\bf S}_i  \cdot {\bf S}_j$.
Here, ${\bf S}_i$ is the spin-$1/2$ operator at site $i$, and $J_1 (=1)$ and $J_2$ are the nearest-neighbor and next-nearest-neighbor couplings, respectively.  
When $J_2$ is finite, the spin configuration cannot satisfy the energy gain by the $J_1$ and $J_2$ interactions simultaneously (frustration). 
Around $J_2=0.5$, where the frustration is strong, an exotic state of matter, quantum spin liquid without any symmetry breaking, might be stabilized as the ground state~\cite{jiang_2012,hu_2013,morita_2015,wang_2018,ferrari_2020,liu_arXiv,nomura_arXiv_1}. 
The model also attracts attention because of its possible relevance to the physics of high-$T_{\rm c}$ cuprates~\cite{Nori_1991,Nori_1992,Nori_1995}.
However, because numerically exact QMC results are not available due to the sign problem, the ground-state phase diagram is still under active debate.

For this challenging problem, the wave functions using neural networks have started to be applied to the zero temperature calculation~\cite{liang_2018,choo_2019,ferrari_2019,westerhout_2020,szabo_2020,nomura_arXiv_1,Nomura_2021}. 
However, to detect a hallmark of the possible quantum spin liquid phase experimentally, the finite-temperature behavior needs to be elucidated. 
Here, we apply Method (II) to perform the finite-temperature calculations for $J_2=0$ and 0.5. 
To check its accuracy, we compare the results with numerically exact ones obtained by Method (I) for the non-frustrated case ($J_2=0$). 
For the frustrated case ($J_2=0.5$), Method (I) suffers from the sign problem, but, up to a $6 \times 6$ lattice, the TPQ results are available, which are also numerically exact.  
\textcolor{black}{Therefore, we perform calculations using the $6 \times 6$ lattice with the total magnetization restricted to be zero $ \bigl (\sum_i S^z_i = 0 \bigr)$.
To further improve the accuracy of the calculation, we utilize the translational and point-group symmetry of the extended system~\cite{Nomura_2021}.  
See SM for the practical details~\cite{Note3}.}

Figure~\ref{fig:J1J2} shows Method (II) results (symbols) for the temperature dependence of the (a) energy, (b) specific heat, and (c) $z$ component of spin structure factor $S^z(\pi,\pi)$, which quantifies the N\'eel-type antiferromagnetic correlation.
We can see that, by the frustration, the antiferromagnetic correlation is largely suppressed, and the entropy release slows down. 
Method (II) results accurately reproduce the exact imaginary-time evolution,
showing its reliability even in the frustrated regime. 

%In Method (II), by optimizing parameters numerically, we obtain a more compact network to represent finite-temperature states compared to the analytically derived network in Method (I): the number of hidden units is $8N_{\rm site}$ in this case, which is compared to a number of ${\mathcal O}(N_\tau N_{\rm site})$ in Method (I). 
%Considering that the scaling of the hidden units is polynomial with respect to the system size, 
%our results are strongly encouraging to expect a computational advantage in even more dedicated simulations for larger systems. 
%\textcolor{black}{Our results show that shallow wide networks are already powerful enough to simulate practical cases with scalable cost, whereas we may alternatively employ deep networks whose formal expressibility is known to be even more powerful~\cite{gao_2017, sharir2021neural}. 
% }

\textcolor{black}{In Method (II), by optimizing parameters numerically, we obtain a more compact and dense network to represent finite-temperature states compared to the analytically derived network in Method (I): the number of hidden units is $N_h = 8N_{\rm site}$ in this case, which is in contrast to  ${\mathcal O}(N_\tau N_{\rm site})$ in Method (I).
The compactness of the network without $d$ spins results in the absence of the negative weights in the MC sampling.
Considering the observation that the $N_h$ scales polynomially with respect to the system size, our results are strongly encouraging to expect a computational advantage in even more dedicated simulations for larger systems. }

\black{
From the application to the Heisenberg model, we see that both the implementation of symmetrization and increasing $N_h$ contributes to enhancing the accuracy of calculations~\cite{Note3}.
It is an important future task to check the reliability of Method (II) for other Hamiltonians, since the convergence of the error in the imaginary-time evolution with respect to $N_h$ should be model dependent. 
}

{\it Summary and outlook.---} 
In summary, we have proposed two cutting-edge approaches that utilize DBMs to simulate the finite-temperature properties of quantum many-body systems.
In the first approach, we provide a deterministic construction of DBMs that exactly represents Gibbs states, which proves the suitability and flexibility of neural networks for encoding thermal properties.
In the second approach, the DBM network parameters are optimized stochastically so that the imaginary time evolution can be approximated efficiently, even for one of the most challenging 2D problems, such as frustrated systems.

Several future directions can be envisioned.
It is an interesting question how the neural-network quantum states perform under other schemes of finite-temperature calculation such as TPQ methods. 
All variational ansatze are, by construction, not powerful enough to express Haar random states, which are taken as the initial states in TPQ calculations. 
Nonetheless, results by tensor-network-based algorithms~\cite{white_2009} imply that using the truncated Hilbert space is sufficient in practical simulations.
The major obstacle is considered to be the entanglement growth along the time propagation, which we expect to be simulated well by neural networks, based on previous works on real-time evolution~\cite{carleo_2017, schmitt_2020, xie_2021}.
Also, it is natural to explore the scalability of our methods in larger and/or more complex systems, or ask whether other network structures \textcolor{black}{(e.g. deep feed-forward networks)} are suited for the finite-temperature calculations;
\textcolor{black}{the trade-off relationship between the representability and trainability of shallow/deep neural networks remains to be open.} 
%question not only for ordinary optimization problems, but also for variational simulations such as real/imaginary-time evolution.}

\section{Acknowledgements}
Y.N. is supported by Grant-in-Aids for Scientific Research (JSPS KAKENHI) (Grants No. 16H06345, No. 17K14336, No. 18H01158, and No. 20K14423) and MEXT as ``Program for Promoting Researches on the Supercomputer Fugaku'' (Basic Science for Emergence and Functionality in Quantum Matter ---Innovative Strongly-Correlated Electron Science by Integration of ``Fugaku'' and Frontier Experiments---) (Project ID: hp200132).
N.Y. is supported by the Japan Science and Technology Agency (JST) (via the Q-LEAP program).
F.N. is supported in part by: NTT Research,
Army Research Office (ARO) (Grant No. W911NF-18-1-0358),
Japan Science and Technology Agency (JST)
(via the CREST Grant No. JPMJCR1676),
JSPS (via the KAKENHI Grant No. JP20H00134
and the JSPS-RFBR Grant No. JPJSBP120194828),
the Asian Office of Aerospace Research and Development (AOARD) (via Grant No. FA2386-20-1-4069),
and the Foundational Questions Institute Fund (FQXi) via Grant No. FQXi-IAF19-06.
Part of numerical calculations were performed using NetKet~\cite{netket}.
Some calculations were performed using the computational resources of supercomputers in RIKEN (Fugaku and HOKUSAI) and the Institute of Solid State Physics at the University of Tokyo.

%\appendix

\appendix

\bibliography{bib_yoshioka}

\pagebreak
\widetext
\begin{center}
	\textbf{\large Supplemental Materials for: Purifying Deep Boltzmann Machines for Thermal Quantum States}
\end{center}

%%%%%%%%%% Prefix a "S" to all equations, figures, tables and reset the counter %%%%%%%%%%
\setcounter{section}{0}
\setcounter{equation}{0}
\setcounter{figure}{0}
\setcounter{table}{0}
\renewcommand{\theequation}{S\arabic{equation}}
\renewcommand{\thefigure}{S\arabic{figure}}
\renewcommand{\thetable}{S\arabic{table}}
% \renewcommand{\bibnumfmt}[1]{[S#1]}
% \renewcommand{\citenumfont}[1]{S#1}
%%%%%%%%%% Prefix a "S" to all equations, figures, tables and reset the counter %%%%%%%%%%

\section{Practical details for Method (I)}\label{app:1}
\subsection{Analytical imaginary time evolution for transverse-field Ising model}\label{app:1_ite}
\textcolor{black}{
In Method (I), we construct a DBM such that the imaginary-time evolution is realized analytically within the Trotter error.
As a concrete example, let us consider the transverse-field Ising (TFI) model on the $N_{\rm site}$ spin chain with periodic boundary condition,
\begin{eqnarray}
 \ham = \ham_1 + \ham_2,
\end{eqnarray}
where ${\mathcal H_1} = - J \sum_i \sigma^z_i \sigma^z_{i+1} $ is the Ising-type interaction with amplitude $J$ and ${\mathcal H_2} = - \Gamma \sum_i \sigma^x_i $ is the transverse magnetic field with amplitude $\Gamma$, where $\sigma_i^a \ (a=x, y, z)$ denotes the Pauli matrix operating on the $i$-th site. 
In the following, we provide the solutions to realize a small-time propagation $\dtau$
by updating the DBM parameters from ${\mathcal W}$ to $\bar{\mathcal W}$, i.e.,  $e^{-\delta_{\tau} \mathcal{H}_{\nu}}  | \Psi_{\mathcal W} \rangle = C | \Psi_{\bar{\mathcal W}} \rangle $ ($C$ is a constant)~\cite{carleo_2018}.
}
\begin{figure}[b]
%\vspace{0cm}
\begin{center}
\includegraphics[width=0.98\textwidth]{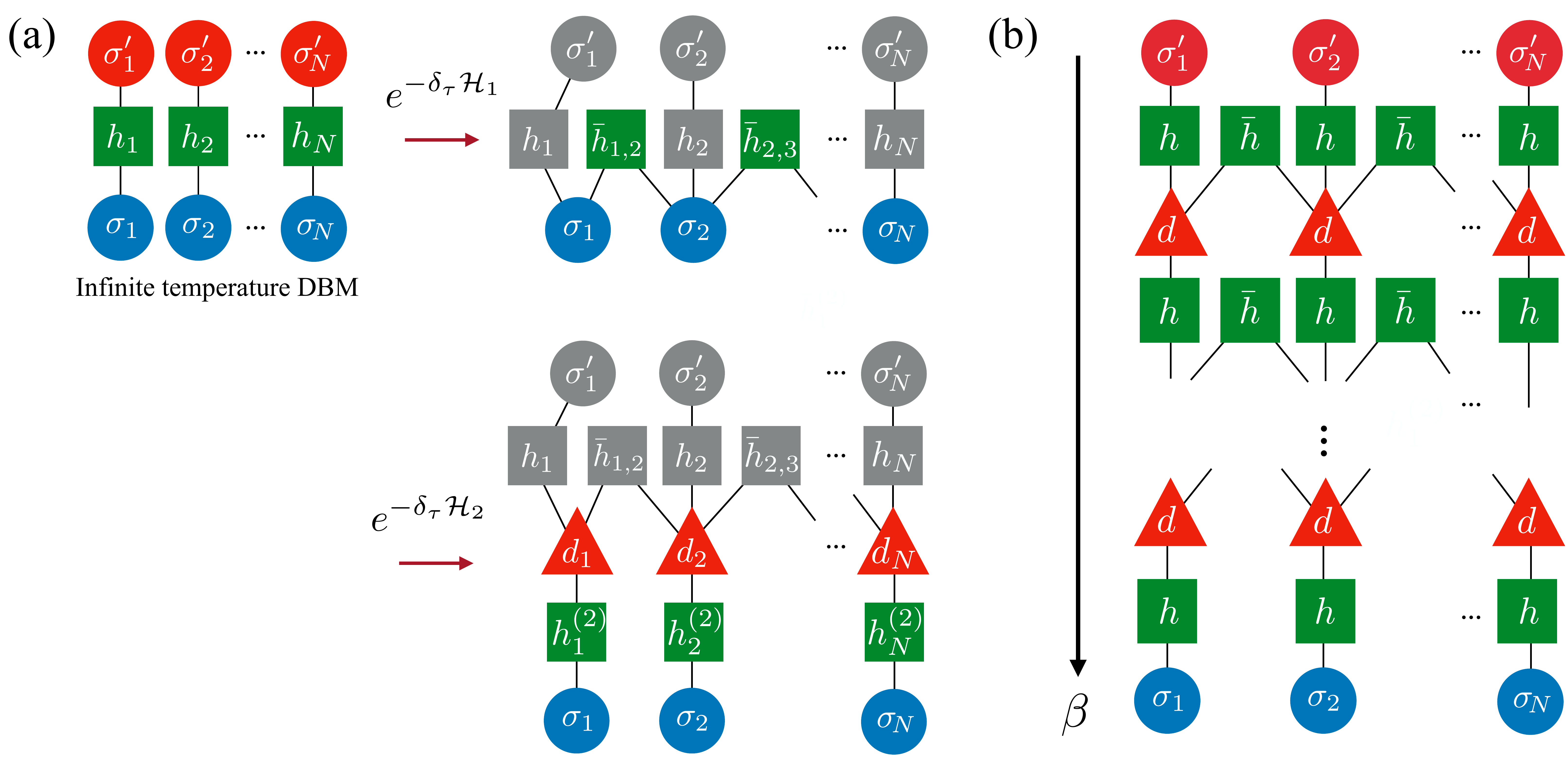}
%\vspace{0cm}
\caption{ \textcolor{black}{Analytical construction of a DBM which realizes the imaginary-time evolution of the  transverse-field Ising model. (a) Starting from the DBM representing the infinite-temperature state, we first encode the interaction propagator ($\nu = 1$) by introducing a single hidden spin for every interacting visible spins. Then, for the magnetic-field propagator ($\nu=2$), we add deep- and hidden-spin layers in between the visible layer and neighboring hidden layer. 
(b) As we repetitively encode the propagators, the network structure grows vertically, with the depth increasing in proportion to the number of Trotter steps.
}
}
\label{fig:TFI_method1}
\end{center}
\end{figure}

\textcolor{black}{
For $\nu=1$ (interaction propagator), a solution is to add one hidden spin per bond and put the couplings with strength $\frac{1}{2}{\rm arcosh} (e^{2 J \delta_\tau })$ to the visible spins on each bond [See Fig~\ref{fig:TFI_method1}(a)].
}

\textcolor{black}{
This is followed by the propagator for $\nu = 2$ (transverse-field propagator), for which we perform the following for each site: (1) cut the existing couplings between the hidden and visible spins and (2) add a new deep spin $d_{\rm new}$ that is coupled with the hidden spins. 
}

\textcolor{black}{
After applying (1) and (2) for each site, we furthermore introduce a new hidden spin which is coupled to the visible and the newly added deep spins $\{d_{\rm new}\}$ with the interaction $\frac{1}{2} {\rm arcosh} \left( \frac{1}{{ \rm tanh} ( \Gamma \delta_\tau )}\right ) $. 
}

\textcolor{black}{
By repeating these operations until the desired inverse temperature is reached, we obtain a DBM that exactly encodes the Gibbs state via the quantum-to-classical correspondence [See Fig~\ref{fig:TFI_method1}(b)]. Here, the number of the hidden and deep spins scales as $O(N_{\tau} N_{\rm site})$, where $N_{\tau}$ is the number of Trotter steps and $N_{\rm site}$ is the number of physical spins.
}

\textcolor{black}{
While we have described the case for the lowest-order Suzuki-Trotter decomposition, we may alternatively consider higher-order ones so that the errors are suppressed.
In the actual numerical implementation for the 1D TFI model, we employ the 2nd-order expansion
\begin{eqnarray}
 {\rm exp}\left[ -\dtau (\ham_1 + \ham_2) \right] &=& {\rm exp} ( -\dtau \ham_1/2) \ \! {\rm exp} (-\dtau \ham_2) \ \! {\rm exp}(-\dtau \ham_1/2) + \mathcal{O}(\dtau ^3).
 %&=& e^{-\frac{7}{24}\dtau \ham_1} e^{-\frac{2}{3}\dtau \ham_2} e^{-\frac{3}{4}\dtau \ham_1}
 %e^{\frac{2}{3}\dtau \ham_2} e^{\frac{1}{24}\dtau \ham_1}e^{-\dtau \ham_2} + \mathcal{O}(\dtau ^4).
\end{eqnarray}
}

\subsection{Monte Carlo sampling scheme for Method (I)}\label{app:1_mc}

Here, we describe how we calculate the expectation value of a physical observable $\langle {\mathcal O} \rangle$ from the constructed DBM states. 
Considering that the norm of the DBM states is given by 
\begin{eqnarray}
\langle \Psi |  \Psi \rangle = \sum_{\sigma, \sigma'} \bigl | \Psi(\sigma,\sigma') \bigr |^2  =  \sum_{\sigma, \sigma'} \sum_{h_1,d_1,h_2,d_2} \phi^{\ast}(\sigma,\sigma'; h_1, d_1) \phi(\sigma,\sigma'; h_2, d_2) =   \sum_{\sigma, \sigma'}  \sum_{h, d} w (\sigma, \sigma'; h,  d )
\end{eqnarray} 
with $ w (\sigma, \sigma'; h,  d ) \equiv   \phi^*(\sigma,  \sigma'; h_1,  d_1) \phi(\sigma, \sigma'; h_2,  d_2)$, the expectation value $\langle {\mathcal O} \rangle$ can be evaluated as 
\begin{eqnarray}
 \langle {\mathcal O} \rangle = \frac{   \langle \Psi( T ) |{\mathcal O} \otimes {\mathbb 1}' | \Psi( T ) \rangle }{ \langle \Psi( T ) |  \Psi( T ) \rangle }
 = \frac{  \sum_{\sigma, \sigma'}  \sum_{h, d} w (\sigma, \sigma'; h,  d ) O_{\rm loc} (\sigma, \sigma'; h,  d ) }{ \sum_{\sigma, \sigma'}  \sum_{h, d} w  (\sigma, \sigma'; h,  d )},
\end{eqnarray}
where the sum over $\sigma$, $\sigma'$, $h$, and $d$ is numerically approximated by MC sampling with weight $ w (\sigma, \sigma'; h,  d ) $.
Here, the ``local'' observable $O_{\rm loc} (\sigma, \sigma'; h,  d )$ reads  
\begin{eqnarray}
 O_{\rm loc} (\sigma, \sigma'; h,  d )  =  \frac{1}{2} \sum_{   \varsigma  } 
 \left( \left\langle \varsigma \right|\mathcal{O}\left|\sigma \right\rangle  \frac{ \phi^*(\varsigma, \sigma'; h_1,  d_1) }{ \phi^*(\sigma, \sigma'; h_1,  d_1) }+ \left\langle \sigma \right|\mathcal{O}\left|\varsigma \right\rangle  \frac{ \phi(\varsigma, \sigma'; h_2,  d_2) }  {\phi(\sigma, \sigma'; h_2,  d_2) } \right),\label{eqn:Oloc}
\end{eqnarray}
where the sum over $\varsigma = (\varsigma_1, ..., \varsigma_{N_{\rm site}})$ is taken over all configurations.
Note that non-diagonal elements $\langle \varsigma| \mathcal{O} | \sigma \rangle$ ($\varsigma \neq \sigma$) are mostly zero when the size of the support of $\mathcal{O}$ is finite. 
For instance, if $\mathcal{O}$ is a product of Pauli operators with the total number of bit-flipping operators (i.e., $\sigma_x$ or $\sigma_y$) given as $k$, the number of non-zero elements contributing in Eq.~\eqref{eqn:Oloc} is $2^k$.

Alternatively, we can use a marginal probability as the weight for the Monte Carlo (MC) method by tracing out one of $h$ and $d$ degrees of freedom.
Here, as an example, we show the case where the $h$ spins are traced out (we can also trace out the $d$ spins and sample the $h$ spins). 
In this case, the marginal probability is given by 
\begin{eqnarray}
  \tilde{w}  (\sigma, \sigma';  d ) \equiv \sum_{h_1,h_2}   \phi^*(\sigma,  \sigma'; h_1,  d_1) \phi(\sigma, \sigma'; h_2,  d_2) =  \tilde{\phi}^*(\sigma,  \sigma';  d_1) \tilde{\phi} (\sigma, \sigma'; d_2), 
\end{eqnarray}
with 
\begin{eqnarray}
 \tilde{\phi} (\sigma, \sigma'; d)   =   \prod_j 2 \cosh  \left [ b_j   +   \sum_i ( W_{ji}  \sigma_i    +  W'_{ji}  \sigma_i^{\prime} )  +   \sum_k W'_{jk}  d_k  \right ].
\end{eqnarray}
The formula for the expectation value $\langle {\mathcal O} \rangle$ is recast as 
\begin{eqnarray}
 \langle {\mathcal O} \rangle = \frac{  \sum_{\sigma, \sigma'}  \sum_{ d} \tilde{w} (\sigma, \sigma';  d ) \tilde{O}_{\rm loc} (\sigma, \sigma';  d ) }{ \sum_{\sigma, \sigma'}  \sum_{d} \tilde{w}  (\sigma, \sigma';  d )}, 
 \label{eq_marginal_O}
\end{eqnarray}
where the local observable $\tilde{O}_{\rm loc} (\sigma, \sigma';  d )$ is given by 
\begin{eqnarray}
 \tilde{O}_{\rm loc} (\sigma, \sigma';   d )  =  \frac{1}{2} \sum_{   \varsigma  } 
 \left( \left\langle \varsigma \right|\mathcal{O}\left|\sigma \right\rangle  \frac{ \tilde{\phi}^*(\varsigma, \sigma'; d_1) }{ \tilde{\phi}^*(\sigma, \sigma';  d_1) }+ \left\langle \sigma \right|\mathcal{O}\left|\varsigma \right\rangle  \frac{ \tilde{\phi} (\varsigma, \sigma';  d_2) }  {\tilde{\phi}(\sigma, \sigma';  d_2) } \right).  
\end{eqnarray}
In the present study, we evaluated the physical observables based on Eq.~(\ref{eq_marginal_O}), since the degrees of freedom participating in the sampling are mitigated.

\section{Practical details of Method (II)}\label{app:2}
\subsection{Stochastic Reconfiguration method}\label{app:2_SR}
\textcolor{black}{The Stochastic Reconfiguration (SR) method~\cite{sorella_1998, sorella_2001} employs a variational principle such that the exact imaginary-time evolution is approximated within the expressive power of the wave function ansatz.
In other words, the SR method provides an update rule of variational parameters so that the distance, whose definition relies on the employed variational principle,  between states following exact and approximate imaginary-time evolution is minimized.
In the following, we provide a concise derivation of the parameter update rule used in the SR method.
}

\textcolor{black}{
Let $\ket{\Psi_{\theta}}$ be a variational wave function with a set of complex variational parameters $\theta$. 
In the SR method, we update the variational parameters as $\theta \leftarrow \theta +\delta \widetilde{\theta}$ according to a variational principle based on the Fubini-Study metric $\mathcal{F}$:
\begin{eqnarray}
 \delta \widetilde{\theta} &=& \mathop{\rm arg~min}\limits_{\delta\theta} \mathcal{F}[e^{-\dtau\mathcal{H}}\ket{\Psi_{\theta}}, \ket{\Psi_{\theta + \delta \theta}}], \\
 \mathcal{F}[\ket{\psi}, \ket{\phi}] :&=& \arccos \sqrt{\frac{\bra{\psi}\ket{\phi} \bra{\phi}\ket{\psi}}{ \bra{\psi}\ket{\psi}\bra{\phi}\ket{\phi} }}.
\end{eqnarray}
Here, $\ham$ is the Hamiltonian of the system and $\dtau$ is a small step of imaginary-time evolution.
Note that the metric $\mathcal{F}$ measures the distance between two quantum states $\ket{\psi} $ and $\ket{\phi}$ after imposing an appropriate normalization,
and therefore it is closely related with the fidelity between pure states $F$ as $\mathcal{F} = \arccos \sqrt{F^2}$, where the fidelity is defined as 
 $F^2[\ket{\psi}, \ket{\phi}] = \frac{\bra{\psi}\ket{\phi} \bra{\phi} \ket{\psi}}{\bra{\psi}\ket{\psi} \bra{\phi} \ket{\phi}}.$
Consequently, the parameter update $\delta \widetilde{\theta}$ also satisfies
\begin{eqnarray}
 \delta \widetilde{\theta} = \mathop{\rm arg~max}\limits_{\delta\theta} F^2[e^{-\dtau \ham}\ket{\Psi_{\theta}}, \ket{\Psi_{\theta + \delta \theta}}].
\end{eqnarray}
}

After some algebraic calculations, we obtain the explicit expression, up to the second order, of the squared fidelity $F^2$ as
\begin{eqnarray}
 F^2 &=& 1 - \left (\sum_{k, l}\delta \theta_k ^* S_{kl} \delta \theta_l  + \dtau \sum_k (f_k \delta \theta_k^* + {\rm c.c.}) + \dtau^2(\langle \ham^2\rangle - \langle \ham \rangle^2) \right), \label{eqn:sqfid}
\end{eqnarray}
where the variance of the energy, contributing to the intrinsic algorithm error, is estimated using the MC sampling as $\langle \cdot \rangle := \frac{\bra{\Psith} \cdot \ket{\Psith}}{\bra{\Psith} \ket{\Psith}}$. [See \ref{app:2_MC} for further information.]
We have introduced an element of the covariance matrix $S_{kl}$ as
\begin{eqnarray}
 S_{kl} &=& \frac{\bra{\partial_k \Psith} \ket{\partial_l \Psith}}{\bra{\Psith} \ket{\Psith}} - \frac{\bra{\partial_k \Psith} \ket{\Psith}}{\bra{\Psith} \ket{\Psith}} \frac{\bra{\Psith} \ket{\partial_l \Psith}}{\bra{\Psith} \ket{\Psith}} \\
 &=& \langle O^{\dagger}_k O_l\rangle - \langle O_k^{\dagger}\rangle \langle O_l \rangle,\label{eqn:smat_def}
\end{eqnarray}
and the generalized force $f_k$ as
\begin{eqnarray}
 f_k &=& \frac{\bra{\partial_k \Psith} \ham \ket{\Psith}}{\bra{\Psith} \ket{\Psith}} - \frac{ \bra{\partial_k \Psith} \ket{\Psith} }{\bra{\Psith} \ket{\Psith}}  \frac{\bra{\Psith} \ham \ket{\Psith}}{\bra{\Psith} \ket{\Psith}} \\
 &=& \langle O^{\dagger}_k \ham \rangle - \langle O_k^{\dagger}\rangle \langle \ham \rangle,\label{eqn:f_def}
\end{eqnarray}
where $O_k$ denotes a diagonal matrix with each element given as $ O_k(\sigma) = \frac{\partial_k\Psith(\sigma)}{\Psith(\sigma)}$. 
Equations.~\eqref{eqn:smat_def} and \eqref{eqn:f_def} indicates that all the elements $S_{kl}$ and $f_k$ can be estimated by MC sampling. 
Finally, by using the stationary condition for the optimal solution of Eq.~\eqref{eqn:sqfid}, the expression for the parameter update can be obtained as
\begin{eqnarray}
\delta \widetilde{\theta}_k =  - \dtau \sum_l S_{kl}^{-1} f_l.
\end{eqnarray}

\textcolor{black}{
In the actual calculation, we add a stabilization factor to the diagonal elements of $S$ as $S_{kk} + \epsilon_k$, where $\epsilon_k$ is typically taken uniformly as $\sim 10^{-4} \bar{S}_{\rm diag}$, where $\bar{S}_{\rm diag}$ is the mean value of diagonal elements.
We observe that the calculation becomes more reliable when the time step $\dtau$ is initially taken to be small; $O(10^{-4})$. After tens to hundreds of iterations, $\dtau$ is increased gradually up to $O(10^{-2})$.
}

\subsection{Symmetrization} \label{app:2_symm}
In Method (II), we numerically optimize a purified DBM wave function with the form: 
\begin{eqnarray}
\Psi(\sigma,\sigma') =  \prod_j 2 \cosh  \Bigl[ b_j \! + \!  \sum_i ( W_{ji}  \sigma_i  \! +\!  W'_{ji}  \sigma_i^{\prime} )  \Bigr ].
\end{eqnarray}
Here, $b$, $W$, and $W'$ are variational parameters. 
We assign the site index $i$ for the pair of physical $\sigma$ and ancilla $\sigma'$ spins.

To improve the quality of the calculation, we consider the symmetry of the system. 
In the case of the 2D $J_1$-$J_2$ Heisenberg model, we encode the translational and point-group symmetry in the purified DBM wave function.
We observe that the initial state, i.e., the purified infinite-temperature state, is invariant under certain symmetry operations: (1) translation $T_{\bf R}$ that shifts all the spins by the amount ${\bf R}$ as $(\sigma, \sigma') \mapsto (T_{\bf R}\sigma, T_{\bf R}\sigma')$, and (2) symmetry operation $R$ of $C_{4v}$ point group that maps a spin configuration as $(\sigma, \sigma') \mapsto (R \sigma, R\sigma')$.
Alternatively, we can understand that the purified state is in the zero wave-number sector and belongs to the $A_1$ representation of the $C_{4v}$ point group (see also Table~\ref{tab_irrep}) of a bilayer system.
Along the imaginary time evolution, the purified DBM wave function stays within the identical symmetry sector at arbitrary temperature.
Hence, we impose such a condition by symmetrizing the purified DBM wave function as 
\begin{eqnarray}
   \Psi_{{\rm sym.}}  ( \sigma , \sigma')   =    \sum_{R, {\bf R} }    \Psi ( T_{\bf R} R  \sigma , T_{\bf R} R \sigma').
   \label{eq_psi_sym}
\end{eqnarray}
We employ this formula to impose the symmetry on the wave function on the left-hand side $\Psi_{\rm sym.}  ( \sigma,  \sigma' )$.
Observe that $\Psi_{\rm sym.}  ( \sigma,  \sigma' )$ satisfies the symmetry even when the bare DBM wave function $\Psi ( \sigma, \sigma' )$ on the right-hand side does not preserve the symmetry. 
We emphasize that the corresponding Gibbs state of the target system  consists of contributions from all symmetry sectors of the original Hamiltonian.

\begin{table}
\caption{Character table of the $C_{4v}$ point group.}
\label{tab_irrep}
\begin{center}
\begin{tabular}{@{\ \ \  } c  |   c @{\ \  \ \  \ }  c @{\ \ \ \ \ } c @{\ \ \ \ \ }  c @{\ \ \ \ \ }   c @{\ \ } }
\hline\hline
  &  \ \ E  & $2C_4$ & $C_2$ & $2\sigma_v$ & $2\sigma_d$  \\
\hline
$A_1$ \  & \ \  1  &  \ \  1  & \ \ 1   & \ \ 1  & \ \ 1  \\
$A_2$ \  & \ \  1  &  \ \ 1   & \ \ 1   & $-1$   & $-1$   \\
$B_1$ \  & \ \  1  & $-1$     & \ \ 1   & \ \  1 & $-1$   \\
$B_2$ \  & \ \  1  & $-1$     & \ \ 1   & $-1$   & \ \ 1  \\
$E$   \  & \ \  2  &  \ \  0  & $-2$    & \ \ 0  & \ \ 0  \\
\hline\hline
\end{tabular}
\end{center}
\end{table}

\subsection{Computing expectation values of physical observables}\label{app:2_MC}

As in Method (I), expectation values are numerically evaluated by the MC method. 
The formula for the expectation value $\langle {\mathcal O} \rangle$ in Method (II) reads
\begin{eqnarray}
 \langle {\mathcal O} \rangle = \frac{  \sum_{\sigma, \sigma'}  p (\sigma, \sigma') O_{\rm loc} (\sigma, \sigma') }{ \sum_{\sigma, \sigma'}  p (\sigma, \sigma' )}, 
\end{eqnarray}
where $p (\sigma, \sigma')$ is the weight $p (\sigma, \sigma') \equiv |\Psi_{\rm sym.} (\sigma,\sigma') |^2 $, and the local observable $O_{\rm loc} (\sigma, \sigma')$ is given by 
\begin{eqnarray}
 O_{\rm loc} (\sigma, \sigma')  =  \frac{1}{2} \sum_{   \varsigma  } 
 \left( \left\langle \sigma \right|\mathcal{O}\left|\varsigma \right\rangle    \frac{  \Psi_{{\rm sym.}}( \varsigma, \sigma') }{ \Psi_{{\rm sym.}}( \sigma, \sigma') } + c.c. \right).  
\end{eqnarray}
We perform the Metropolis sampling over the $\sigma$ and $\sigma'$ spins with the weight $p (\sigma, \sigma')$ to compute expectation values.

\subsection{Calculation conditions} 
\label{sec_calc_condition}

In the present calculations for the 2D $J_1$--$J_2$ Heisenberg model on the $6 \times 6$ lattice (Figure 3 in the paper), we introduce 8$N_{\rm site}$ (=288) hidden spins. 
We set the $b$ parameters (magnetic field) to zero, so that the ``up" and ``down" spins are equivalent, and optimize only the $W$ and $W'$ parameters.
With this setting, the purified DBM wave function becomes even with respect to the global spin inversion.

The initial $W$ and $W'$ parameters are prepared to represent the infinite-temperature state. 
The infinite-temperature state can be reproduced exactly by setting $W_{ji} = W'_{ji} = \mathrm{i} \frac{\pi}{4} \delta_{ji} $, for $1 \leq j  \leq N_{\rm site}$, and $W_{ji} = W'_{ji} = 0 $, for $N_{\rm site}+1 \leq j  \leq 8N_{\rm site}$. 
In the actual calculations, to make the initial gradient of the parameter optimization finite, we put small perturbations to the above setting by adding small random numbers.

As we describe in the main text, starting from the initial $W$ and $W'$ values, we optimize the parameters with the SR method~\cite{sorella_2001}, which makes it possible to reproduce the imaginary time evolution as much as possible within the representability of the DBM. 
To reduce the number of variational parameters, we take half of the $W$ and $W'$ parameters to be complex and the rest real as in Ref.~\onlinecite{Nomura_2021}.
We do not impose symmetry constraints on the $W$ and $W'$ parameters, but instead, the symmetry is restored with the projection [Eq.~(\ref{eq_psi_sym}) in the case of the 2D $J_1$--$J_2$ Heisenberg model].
We observe that the most time-consuming part of the calculation is the MC sampling to estimate the expectation values of the energy and the gradient of the parameter optimization; 
its time scales as ${\mathcal O}( N_h N_{\rm site}^2)$, where $N_h$ is the number of hidden spins ($N_h = 8 N_{\rm site}$ in the present case).

\subsection{\black{Benchmark calculations for 1D and 2D Heisenberg models}}

\begin{figure*}[tb]
%\vspace{0cm}
\begin{center}
\includegraphics[width=0.96\textwidth]{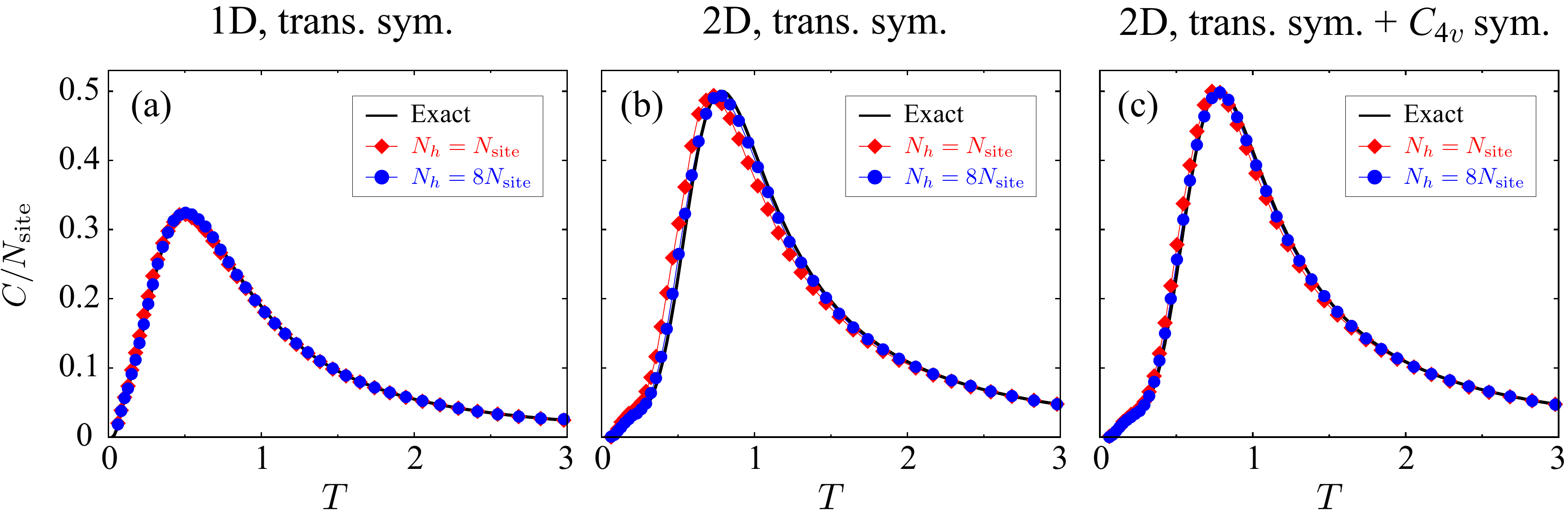}
%\vspace{0.0cm}
\caption{
\black{
Method (II) results (symbols) for the specific heat $C$ for (a) the 1D Heisenberg model on the 16-site chain and (b,c) the 2D Heisenberg model on the 4x4 lattice.
In all cases, we take the periodic boundary condition, and the number of sites is 16 ($N_{\rm site}=16$). 
In (a) and (b), we impose translational symmetry on the purified DBM wave function for the extended system. 
In (c), we use both translational and $C_{4v}$ point-group symmetry.
The solid black curves are obtained by exact diagonalization. }
}
\label{fig:Heis_1D_2D}
\end{center}
\end{figure*}

\black{
Here, we show the benchmark results of Method (II) for the 1D and 2D Heisenberg models. 
The Hamiltonian reads
${\mathcal H} =  J_1 \sum_{ \langle i, j \rangle}  {\bf S}_i  \cdot {\bf S}_j $, 
where $J_1=1$ and $\langle i, j \rangle$ denotes a pair of neighboring sites. 
In the 2D case, the Hamiltonian corresponds to the $J_2=0$ case for the 2D $J_1$--$J_2$ Heisenberg model. 
We take the 16-site chain for the 1D model, and the $4 \times 4$ square lattice for the 2D model. 
In both cases, periodic boundary conditions are assumed and the number of sites is 16 ($N_{\rm site}=16$). 
We consider the zero magnetization sector ($\sum_i S_i^z=0$) as in the main text. 
}

\black{
As we have discussed above and in the main text, the quality of the Method (II) calculations can be improved by increasing the number of hidden units $N_h$ and/or by imposing symmetry.  
For a numerical demonstration, we compare $N_h = N_{\rm site}$ and $N_h = 8N_{\rm site}$ cases. 
We follow the conditions described above (Sec.~\ref{sec_calc_condition}), except that all the parameters are taken to be complex in the $N_h = N_{\rm site}$ case (note that when $N_h = N_{\rm site}$, all the hidden units need to have complex couplings to prepare infinite-temperature states).
}

\black{
Figure~\ref{fig:Heis_1D_2D} shows the Method (II) results for the specific heat $C$ for (a) the 1D model when imposing translational symmetry, (b) the 2D model when imposing translational symmetry, and (c) the 2D model when imposing translational and point-group symmetries. 
We see that in the case of the 1D model, the Method (II) result agrees well with the exact result already at $N_h = N_{\rm site}$ [Figure~\ref{fig:Heis_1D_2D}(a)]. 
On the other hand, in the case of the 2D model, when only the translational symmetry is imposed as in the 1D model case, the result with $N_h = N_{\rm site}$ shows a visible (but small) deviation from the exact result  [Figure~\ref{fig:Heis_1D_2D}(b)]. 
The result improves by increasing $N_h$ [Figure~\ref{fig:Heis_1D_2D}(b)] or imposing an additional symmetry (point-group symmetry) [Figure~\ref{fig:Heis_1D_2D}(c)]. 
From this benchmark, we indeed observe that the number of hidden units $N_h$ and the symmetrization are important factors to ensure the quality of the Method (II) calculations. 
}

\end{document}